\documentclass[12pt]{article}
\usepackage[hypertex]{hyperref}
\usepackage[dvipdfmx]{graphicx} 
\usepackage{amsmath,amssymb,amsfonts,bm,color,cite}

\def\a{\alpha} \def\b{\beta} \def\g{\gamma} \def\G{\Gamma} \def\d{\delta} \def\D{\Delta} \def\e{\epsilon}       \def\m{\mu} \def\n{\nu} \def\x{\xi}     \def\s{\sigma} \def\S{\Sigma}    \def\c{\chi}  

\def\d{{\bm d}}  \def\D{{\bm D}} \def\A{{\bm A}} \def\F{{\bm F}} 

\def\dg{\dagger}
\def\del{\partial}
\newcommand{\we}{\wedge}
\newcommand{\To}{\Rightarrow}

\newcommand{\getsto}{\leftrightarrow}

\newcommand{\tr}{{\rm tr}}
\newcommand{\vev}[1]{ \langle {#1} \rangle }
\def\nn{\nonumber}

\newcommand{\Lg}{\mathcal{L}}

\begin{document}

\begin{titlepage}

\begin{flushright}
KEK-TH-1835
\end{flushright}

\vskip 1.35cm

\begin{center}
{\large \bf SU(5) orbifold GUT in noncommutative geometry}

\vskip 1.2cm

Masaki J.S. Yang

\vskip 0.4cm

{\it Institute of Particle and Nuclear Studies,\\
High Energy Accelerator Research Organization (KEK)\\
Tsukuba 305-0801, Japan\\
}

\date{\today}

\begin{abstract} 
In this paper, we appliy the orbifold GUT mechanism to the SU(5) model in noncommutative geometry, 
including the fermonic sector. 
Imposing proper parity assignments for ``constituent fields'' of bosons and fermions, 
the couplings between fermions and the heavy bosons $X_{\m}, Y_{\m}$ , and $ H^{c}$ 
are prohibited by the parity symmetry. 
As a result, the derived fermionic Lagrangian is just that of the standard model, and proton decay is forbidden at tree level.
If quantum fluctuation respects the parity symmetry, 
the process will be naturally suppressed or even forbidden completely.  

\end{abstract} 

\end{center}
\end{titlepage}

\section{Introduction}

The grand unified theory (GUT) \cite{Georgi:1974sy, Pati:1974yy} is one of the most attractive candidates  
beyond the standard model(SM). This concept is widely applied to other regions, 
supersymmetry \cite{Dimopoulos:1981zb,Sakai:1981gr}, and family unification \cite{Ramond:1979py,Wilczek:1981iz,Kugo:1983ai}, with or without an extra dimension \cite{Raby:2008gh, Babu:2002ti}.

GUT is also applied in the Higgs mechanism inspired noncommutative geometry (NCG) 
\cite{Connes:1990qp, Connes:1994yd, DuboisViolette:1988ir, Coquereaux:1990ev, Sitarz:1993zf, Morita:1993zv}. 
Chamseddine, Felder, and Fr\"olich proposed an SU(5) GUT model in NCG 
\cite{Chamseddine:1992kv, Chamseddine:1992nx}. 
In this context, the underlying spacetime is considered to be product of Minkowski spacetime and discrete points, $M^{4} \times Z_{n}$. The Higgs boson is regarded as a gauge boson between discrete points that has noncommutative differential algebra. 
An advantage of this application is that the couplings of the Higgs sector 
are tightly determined from noncommutativity and the compositeness-like formulation.
By contrast, a shortcoming is that quantum theory is not established completely.
The original paper has followed by several authors \cite{Morita:1993xj, Okumura:1994ck, Sogami:1996xy, Hashimoto:1999qi,Konisi:1998ur}, and extended to an SO(10) model by the original authors \cite{Chamseddine:1993is}.

Meanwhile, when a model in this context is interpreted as a theory with an extra dimension \cite{Lizzi:2000bc, Sarrazin:2009ea}, 
several mechanisms in the usual extra dimension can be diverted to the models in NCG, 
such as the (de)construction \cite{Alishahiha:2001nb}. 
Based on this idea, in the previous study, we applied the orbifold GUT mechanism \cite{Kawamura:2000ev} to the SU(5) GUT in NCG \cite{Yang:2015gsa}. 
However, the application remained in only the bosonic sector. 
Therefore, in this paper, we apply the orbifold GUT mechanism to the SU(5) model in NCG, including the fermonic sector. 
This study corresponds to Refs. \cite{Hebecker:2002rc, Altarelli:2001qj} in the usual orbifold GUT theories. 
In order to achieve a correct breaking scheme of SU(5), 
the background spacetime is assumed to be $M^{4} \times Z_{3}$. 
Imposing proper parity assignments for ``constituent fields'' of bosons and fermions, 
the couplings between fermions and the heavy bosons $X_{\m}, Y_{\m}$, and $ H^{c}$ 
are prohibited by the parity symmetry. 
As a result, the derived fermionic Lagrangian is just that of the SM, and proton decay is forbidden at tree level.
If quantum fluctuation respects the parity symmetry, 
the process will be naturally suppressed or even forbidden completely.  

Moreover, the application of the orbifold GUT mechanism to the fermion sector may be meaningful for 
model building. 
In early papers of the NCG \cite{Coquereaux:1990ev, Morita:1993zv, Chamseddine:1992kv, Chamseddine:1992nx}, 
the {\it ad hoc} chiral condition is usually imposed on the fermions 
in order to produce chiral Yukawa couplings. 
However, if we consider the NCG model as a model with an extra dimension, 
especially with several branes \cite{Hashimoto:1999qi}, 
these chiral fermions will generate a brane anomaly. 
This is a mirror fermion problem \cite{Maalampi:1988va}. 
By the orbifold GUT mechanism, we can eliminate mirror fermions in the proper situation. 

This enables us to implement a new kind of theory in NCG: family unification in the extra dimension \cite{Babu:2002ti}.

This paper is organized as follows. 
In the next section, we review the basic formulation of generalized gauge theory in NCG.
In Sect.~3, the SU(5) model and orbifold GUT mechanism are presented.
Section~4 is devoted to conclusions.

\section{Generalized gauge theory on $M^{4} \times Z_{N}$ }

In this section, we present a basic formulation of generalized gauge theory on $M^{4} \times Z_{N}$. 
The original papers utilize the Dirac operator and the Clifford algebra \cite{Connes:1990qp, Chamseddine:1992kv, Chamseddine:1992nx}. However, we follow the formulation using the inner product of the differential forms, 
 developed in \cite{Morita:1993xj,Okumura:1994ck}.
The formulation presented here is quoted from Ref. \cite{Morita:1993xj,Okumura:1994ck}, 
and is basically the same.

\subsection{Differential calculus and generalized gauge field}

The background spacetime $M^{4} \times Z_{n}$ is the direct product of 
the ordinary four-dimensional Minkowski space $M^{4}$ and discrete space $Z_{n}$, 
with the coordinates $(x^{\mu}, n = 1-N)$. 
In this space, the generalized exterior derivative $\d$ is defined as follows;
\begin{align*}
\d f (x,n) & =(d + d_{\chi}) f (x,n), ~~~~
 d \hspace{0.5pt} f(x,n)  = \partial_\mu f(x,n) dx^\mu, \\
 d_{\chi} f(x,n) &= \sum_{m \neq n} d_{\chi_{m} } f(x,n) = \sum_{m \neq n} [ M_{nm} f (x,m)  - f (x,n) M_{nm}] \chi_{m},  
\end{align*}
where $\chi_{k}^{\dg} = -\chi_{k}$ and $M_{nm}^{\dagger} = M_{mn} ~ (n \neq m)$ are assumed. 
These matrices determine the distance between two Minkowski spacetimes and the pattern of the spontaneous symmetry-breaking. 

In order to preserve the usual Leibniz rule for the extra derivative $d_{\chi}$,
\begin{equation}
d_{\chi_{m}} [f(x,n) g (x,n)]  = [(d_{\chi_{m}} f(x, n)) g(x,n) + f(x,n) (d_{\chi_{m}} g(x,n))]  ~~ (n\neq m) ,
\end{equation}
we should assume the following  ``index shifting rule'': 
\begin{equation}
f (x,m) \, \chi_{m} \, g (x,n) = f (x,m) \, g (x,m) \, \chi_{n} .
\end{equation}
This is the source of the noncommutativity that corresponds to the relation $y \, dy = - dy \, y $ in other formulations \cite{DuboisViolette:1988ir, Coquereaux:1990ev, Sitarz:1993zf, Morita:1993zv}.
For the ``one form'' $M_{nm} \chi_{m} f(x,m)$, the Leibniz rule is modified to be
\begin{equation}
d_{\chi_{l} } [M_{nm} \chi_{m} f(x,m)] = (d_{\chi_{l}} M_{nm} \chi_{m}) f(x,m) - M_{nm} \chi_{m} \we (d_{\chi_{l}} f(x,m) ) ,
\label{graded}
\end{equation}
where 
\begin{align}
d_{\c_{l}} M_{nm} &= M_{nm} M_{ml} \chi_{l}  .
\end{align}
Equation (\ref{graded}) corresponds to the usual graded Leibniz rule of the differential form,
\begin{align}
d(\x \we \c) = d\x \we \eta + (-1)^{\del \x} \x \we d \c ,
\end{align}
where $\del \x$ is the order of the differential form $\x$.

Utilizing this calculus, we can prove the nilpotency of $\d$ that is indispensable to constructing the gauge theory.
By the definition of $\d = d + d_{\chi}$, the nilpotency condition is rewritten as
\begin{align}
\d^{2} f(x,n) = [d^{2} + d d_{\chi} + d_{\chi} d + d_{\chi}^{2}] f(x,n) = 0 . 
\end{align}
Enforcing the ordinary anticommutative relation $dx^{\mu} \we \chi_{m} = - \chi_{n} \we dx^{\mu}$, 
the condition is reduced to 
\begin{align}
d_{\chi}^{2} f(x,n) = 0 .
\end{align}
Here, we impose $\chi_{m} \we \chi_{k} = + \chi_{k} \we \chi_{m}$ due to 
the noncommutative property of the background spacetime. Nevertheless, 
\begin{align}
d_{\chi}^{2} f(x,n) &= d_{\chi} \sum_{m} [ M_{nm} f (x,m) - f(x,n) M_{nm}] \chi_{m} \\
&=  \sum_{m,l} \bigg[ M_{nm} M_{ml} \chi_{l} f (x,m) 
- M_{nm} [ M_{ml} f (x,l) - f(x,m) M_{ml}] \chi_{l}\\
& - [ M_{nl} f (x,l) - f(x,n) M_{nl}] \chi_{l} M_{nm} - f(x,n) M_{nm} M_{ml} \chi_{l} \bigg] \we \chi_{m} \\
&=  \sum_{m,l} 
 \bigg[ + M_{nm} f (x,m) M_{ml}  - M_{nl} f (x,l)  M_{lm} \bigg] \chi_{l} \we \chi_{m} = 0 .
\end{align}
In the last line, another index shifting rule,  
$F_{k n} \chi_{n} M_{ml} = F_{k n} M_{nl} \chi_{n} $\cite{Morita:1993xj}, is applied.
As a result, 
\begin{align}
\d^{2} f(x,n) = 0. 
\end{align}
The proof of the nilpotency in the general case is also presented in \cite{Okumura:1994ck}. 

\vspace{12pt}

Next, we consider the generalized gauge field $\A (x,n)$ in this space:
\begin{align}
\A (x,n) &\equiv \sum_{i} a_{i}^{\dagger} (x,n) \d a_{i} (x,n) \equiv A(x,n) + \sum_{m\neq n} \Phi_{nm} (x) \chi_{m} .
\label{ga}
\end{align}
Here, $a_{i} (x,n)$ is square-matrix-valued function and the summation over $i$ is assumed to be a finite sum. 
In the components, the gauge and Higgs fields are represented as
\begin{align}
A(x,n) &= \sum_{i} a_{i}^{\dagger} (x,n) da_{i} (x,n) , \\
\Phi_{nm} (x) &= \sum_{i} a_{i}^{\dagger} (x,n) [M_{nm} a_{i} (x,m) - a_{i} (x,n) M_{nm}] ~~~ (n\neq m).
\label{Phi}
\end{align}
According to  \cite{Chamseddine:1992kv, Chamseddine:1992nx, Morita:1993xj},  we impose the normalization  condition $\sum_{i} a^{\dagger}_{i} (x,n) a_{i} (x,n) = 1$, which leads to the following Hermitian condition:
\begin{align}
A^{\dagger}(x,n) = -A(x,n), ~~~\Phi_{nm}^{\dagger} (x) = \Phi_{mn} (x) .
\label{Hermitian}
\end{align}

The gauge transformation property of the $a_{i} (x,n)$ is assigned to a fundamental representation under the $n$th gauge transformation,
\begin{equation}
a^{g}_{i} (x,n) = a_{i} (x,n) g (x,n) ,
\label{transai}
\end{equation}
where $g(x,n) = g^{-1} (x,n)^{\dagger}$ is an arbitrary unitary matrix associated with the gauge group on the $n$th $M^{4}$ space. 
From Eq.~(\ref{transai}),  the gauge transformation of $\A (x,n)$ is derived as the standard form 
\begin{align}
\A^{g} (x,n) &= g^{-1} (x,n) \A(x,n) g (x,n) + g^{-1} (x,n) \d g (x,n) , \label{transA}
\end{align}
with $M_{nm}^{'} = M_{nm}$. In particular, the following {\it back-shifted} Higgs field, 
\begin{equation}
H_{nm} (x) \equiv \Phi_{nm} (x) + M_{nm} = \sum_{i} a^{ \dagger}_{i} (x,n) M_{nm} a_{i} (x,m) ,
\label{backshift}
\end{equation}
transforms as a bifundamental representation,
\begin{align}
H_{nm}^{g} (x) = g^{-1}(x,n) H_{nm} (x) g(x,m) ,
\end{align}
and is identified as a physical Higgs boson with a vacuum expectation value.

Regarding this connection $\A$ as a building block, 
 we can construct  the field-strength two-form $\F$, 
\begin{align}
\F(x,n) = \d \A (x,n) + \A (x,n) \we \A (x,n) , ~~~  \d \A = \sum_{i} \d a^{\dg}_{i}(x,n) a_{i} (x,n) ,
\end{align}
and the gauge-invariant Lagrangian 
\begin{equation}
\Lg_{\rm YMH} = - {1\over 4} \sum_{n} {1\over g_{n}^{2}} \tr \vev{\F (x,n) , \F(x,n) } .
\label{LYMH}
\end{equation}
Here, $g_{n}$ are independent coupling constants introduced on each $n$th space. 
The Lagrangian (\ref{LYMH}) is subdivided into four terms:  
The first term is the pure Yang--Mills term with independent coupling constants, 
the second is the Higgs kinetic energy term, 
the third represents the self-coupling of Higgs $H_{nm}$, 
and the fourth term describes interactions among different Higgs, $H_{nm}$ and $H_{ml}$. 
The derivation and explicit formula of Eq.~(\ref{LYMH}) is in Ref.~\cite{Morita:1993xj,Okumura:1994ck}.

\subsection{Fermionic Lagrangian}

Next, we proceed to the fermion sector to construct the full Lagrangian. 
At first, we introduce the {\it generalized spinor one form} $\D \psi$, and the covariant derivative $\D$ acting on the spinor field $ \psi (x,n) $
by
\begin{equation}
\D \psi (x,n) = ( \d + \A^f(x,n)) \psi (x,n).
\end{equation}
Here, $\A^f(x,n)$
is the differential representation for the fermions $\psi (x,n)$ such that
\begin{equation}
\A^f(x,n)=A_{\mu}^f (x,n) dx^\mu+\sum_m \Phi^{f}_{nm} (x) \chi_m.
\end{equation}
Note that $A_{\mu}^f (x,n) (\Phi_{nm}^{f} (x))$ does not necessarily agree with boson
$A_{\mu} (x,n) (\Phi_{nm} (x))$ in the $n$th space of $M^{4}$. 

We also define the extra derivative of the fermion as
\begin{align}
d_\chi \psi (x,n)&=\sum_m d_{\chi_m} \psi (x,n) 
=\sum_m M_{nm}^{f} \chi_m \psi (x,n) = \sum_m M_{nm}^{f} \psi (x,m) \chi_m ,
\label{dchi}
\end{align}
which leads to
\begin{equation}
\D \psi (x,n) =  [(\del_\mu + A_{\mu}^f (x,n) ) dx^\mu
+\sum_m H^f_{nm} (x) \chi_m ] \psi (x,n) .
\label{covpsi}
\end{equation}
Here we used $H^f_{nm}(x)=\Phi^f_{nm}(x)+M^f_{nm}$, and $M_{nm}^f$ in Eq.~(\ref{dchi})
is the corresponding expression to $\Phi_{nm}^f$.

Henceforth we investigate the gauge transformation property
of $ \D  \psi (x,n) $.
The gauge transformation of $ \psi (x,n) $ is defined to be
\begin{equation}
   \psi (x,n)' = [g^f (x,n)]^{-1}  \psi (x,n) , \label{psiprime}
\end{equation}
where $g^f (x,n)$ is the gauge transformation function corresponding
to the representation of $ \psi (x,n) $.
Due to this, $A_{\mu}^f (x,n)$ and $H^f_{nm}$ should transform as
\begin{align}
  {A_{\mu}^f (x,n)}' &= [g^f (x,n)]^{-1}dg^f (x,n) + [g^f (x,n)]^{-1} A_{\mu}^f (x,n) g^f (x,n), \label{Aprime} \\
 {H_{nm}^f}' &= [g^f (x,n)]^{-1} H_{nm}^{f} g^{f} (x,m) .   \label{Hprime}
\end{align}
From Eqs.~(\ref{psiprime}), (\ref{Aprime}), and (\ref{Hprime}), we can easily verify that ${\D} \psi (x,n) $ is gauge covariant:
\begin{equation}
      \D  \psi'_{n}=(g^f (x,n))^{-1} \D \psi (x,n) . 
\end{equation}

In order to obtain the Dirac Lagrangian by the inner products of differential forms, 
the original paper introduce the following {\it associated spinor one-form} \cite{Okumura:1994ck}:
\begin{equation}
{\tilde  \D } \psi (x,n) = \gamma_\mu  \psi (x,n) dx^\mu - i c_{Y} \psi (x,n) \sum_m\chi_m .
\label{asso}
\end{equation}
Here, $c_{Y}$ is a real, dimensionless constant which relates to the Yukawa coupling constant.
It is obvious that ${\tilde  \D } \psi (x,n) $ is also gauge covariant, 
\begin{equation}
      {\tilde  \D }\psi (x,n)' =(g^f (x,n))^{-1} {\tilde {\D}} \psi (x,n). 
\end{equation}
Finally, we introduce the inner products for spinor one-forms, 
\begin{align}
\vev {A_{n} dx^\mu, B_{m} dx^\nu} &= \bar{A}_{n} B_{m} g^{\mu\nu},\\
\vev {A_{n} \chi_k, B_{m} \chi_l }& =- \bar{A}_{n} B_{m} \alpha^2\delta_{kl}.
\end{align}
Here, $\bar{A}_{n} =A^{\dg}_{n} \gamma^0$ denotes 
the usual Lorentz conjugate of the spinors, while other inner products vanish.

Summarizing these considerations, the
 Lorentz and gauge-invariant Dirac lagrangian is constructed by taking the inner product and the summation over $n=1 - N$:
\begin{align}
{\cal L}_{ D} =& \sum_{n=1}^{N} i \vev
{{\tilde  \D } \psi (x,n) , \D  \psi (x,n) }  \\
=&\sum_{n,m=1}^{N} \bar \psi (x,n) [ i \g^\m (\del_\m+A^{f}_{\mu} (x,n)) \delta_{nm} - c_{Y} \a^{2}  H^f_{nm} (x) ] \psi (x,m) .
\label{DiracL}
\end{align}
In particular, the last term of Eq.~(\ref{DiracL}) provides the 
Yukawa couplings constant $y = c_{Y} \a^{2} $ between Higgs and fermions.

\section{SU(5) grand unified theory}

In this section, we review an SU(5) GUT in the NCG,
 and implement the orbifold GUT mechanism for the fermonic sector.
Since the SU(5) GUT model has two symmetry-breaking scales, 
the model requires $N \geqq 3$, which realizes more than two independent $M_{nm}$s.
Then we choose $N=3$ to construct the SU(5) GUT \cite{Chamseddine:1992kv, Morita:1993xj}. 
The indices $n, m, l$ run the values 1, 2, 3 only.

At the beginning,  $a_{i}(x,(1,2))$ are assumed to be complex $5 \times 5$ matrices and $a_{i}(x,3)$ is a real-valued continuous function that satisfies Eq.~(\ref{Hermitian}), $\sum_{i} a^{\dg}_{i} (x,n) a_{i} (x,n) = 1$. 
Moreover, a parity symmetry between $n = 1,2$ is imposed with 
the following parity condition for the fields:
\begin{equation}
 a_{i}(x,1) = P a_{i} (x,2) P,
 \label{PP}
\end{equation}
where $P$ = diag $(-1,-1,-1,+1,+1)$.
In order to break the gauge symmetry, this parity assignment is found to be unique under proper assumptions \cite{Yang:2015gsa}.

The SU(5) gauge fields at each discrete point are calculated from the $a_{i}(x,n)$s as
\begin{align}
A(x,1) &= \sum_{i} a_{i}^{\dg} (x,1) d  a_{i}(x,1) =  i T^{a} A_{1}^{a} (x) \equiv A,  \\
A(x,2) &= \sum_{i} a_{i}^{\dg} (x,2) d  a_{i}(x,2) = i T^{a} A_{2}^{a} (x) \equiv P A P, \label{A2}\\
A(x,3) &= \sum_{i} a_{i}^{\dg} (x,3) d  a_{i}(x,3) = 0,
\end{align}
where $T^{a} (a=1, \cdots , 24)$ are the generators of SU(5). 
In order to eliminate the redundant $U(1)$ generator, the following traceless condition is imposed:
\begin{equation}
{\rm Tr} A(x,1) = {\rm Tr} A(x,2) = 0.
\end{equation}
The matrix $M_{nm}$ are fixed on as
\begin{align}
M_{12} &= M_{21} = M {\rm diag} (1,1,1,1,1) \equiv \S_{0} , \\
M_{13} &= M_{23} = M_{31}^{\dg} = M_{32}^{\dg} = 
\m \begin{pmatrix} 0 & 0 & 0 & 0 & 1 \end{pmatrix}^{T} \equiv H_{0},
\end{align}
where $M (\m)$ corresponds to  the energy scale of GUT (SM) symmetry-breaking. 
These $M_{nm}$s determine the following back-shifted Higgs fields:
\begin{align}
\S (x) + \S_{0} &= H_{12} (x) = \sum_{i} a_{1}^{i \, \dg} M P a^{i}_{1} P = P H_{21} (x) P,  \label{H12} \\
H(x) + H_{0} &= H_{13} (x) = \sum_{i} a_{1}^{i \, \dg} M_{13} a^{i}_{3} =  P H_{23} (x) . \label{H13}
\end{align}
Here, the field $H_{13}(x) (H_{12}(x))$ is a  $5 \times 1$ ($5 \times 5$) matrix transforming like the 5 (1 plus 24)  representation under SU(5). 

Substituting these results into Eq.~(\ref{LYMH}), it is found that the Lagrangian 
contains the following mass term of the 5 representation Higgs \cite{Yang:2015gsa},
\begin{equation}
\Lg \ni | (MP - M) H |^{2} = M^{2} {\rm diag}(4,4,4,0,0) \, H^{\dg} H ,
\label{higgsmass}
\end{equation}
and the gauge boson masses
\begin{align}
\Lg \ni| D_{\m} H_{12} |^{2} &\ni  (A_{\m} M - M P A_{\m} P)^{2} = (M A_{\m}^{\check a} T^{\check a})^{2} .
\label{gaugemass}
\end{align}
Here, $a = \hat a + \check a, $ $\hat a$ runs the generators of SU(3)$_{c}$ $\times$ SU(2)$_{L}$ $\times$ U(1)$_{Y}$, and 
$\check a$ runs the broken generator except for those of SU(3)$_{c}$ $\times$ SU(2)$_{L}$ $\times$ U(1)$_{Y}$.
Equations~(\ref{higgsmass}) and (\ref{gaugemass}) show that the parity assignment condition Eq.~(\ref{PP}) $a^{i}_{2} = P a^{i}_{1} P$ invokes SU(5) symmetry-breaking, and provides the colored triplet Higgs and broken gauge bosons with heavy mass of order $M$. 
Therefore, it is adequate to regard that this symmetry-breaking by the condition (\ref{PP}) corresponds to the orbifold GUT mechanism \cite{Kawamura:2000ev} of the GUT in NCG.

\subsection{Fermionic sector}

Under the $Z_{2}$ parity symmetry ${1 \getsto 2}, $ the $5$ and $10$ representation fermions in the SU(5) model are assigned at each point as follows:
\begin{align}
   \psi (x,1) = \psi(x,2) = {a_1\over \sqrt{2}} \psi_{10},  ~~~ \psi(x,3) = \psi_{5},
\label{assign}
\end{align}
where $a_1 / \sqrt{2}$ is the normalization coefficient for the final expression, 
and only one generation is assumed for simplicity.
In particular, Eq.~(\ref{assign}) indicates that $\psi_{10}$ is assigned to even charge under the parity symmetry.
In components, the fermions $\psi_{10}, \psi_{5}$ are represented as 
\begin{equation}
   \psi_{10 }^{ij} =
\begin{pmatrix}
0 & u_3^c &-u_2^c & u_1 & d_1 \\
-u_3^c & 0 & u_1^c & u_2 &d_2  \\
u_2^c  & -u_1^c  & 0 & u_3 &d_3\\
-u_1 &-u_2 &-u_3 & 0 & e^c     \\
-d_1 &-d_2 & -d_3 & -e^c  &0     \\
\end{pmatrix}_{L} , ~~~
\psi_{5 }^{i} =  \begin{pmatrix} d_1 \\  d_2 \\  d_3 \\  e^c \\ \n^c\\  \end{pmatrix}_{R} .
   \label{chiral}
\end{equation} 
The subscripts $L,R$ denote that they are chiral fermions. 
These are the ad hoc chiral conditions explained in the introduction.
Utilizing the orbifold mechanism,  if a five-dimensional theory has a vector representation $\psi_{L,R}$, 
there remain only chiral fermions in a low energy four-dimension theory.
However, in this paper, 
the parity assignment is spent to break SU(5) symmetry, 
and then we retain the chiral condition Eq.~(\ref{chiral}).

Each $\psi (x, n)$ transforms under the gauge transformation respectively as
\begin{align}
\psi^{g} (x,(1,2)) &=g(x,(1,2)) \otimes g(x (1,2)) \, \psi(x,(1,2)) ,  \\
\psi^{g} (x,3) &= [g(x,1) + g(x,2)] \psi(x,3),  \label{gaugetrf3}
\end{align}
where $g(x,n)$ is the gauge transformation function belonging to SU(5). 
In fact, Eq.~(\ref{gaugetrf3}) seems to be an {\it ad hoc} condition.  
This point will be discussed in the next subsection. 
With this in mind, 
the generalized spinor one-form $ \D  \psi (x,n) $ in Eq.~(\ref{covpsi}) is taken to be
\begin{align}
 \D \psi(x,1) &=  {a_1\over \sqrt{2}} [ \del_\m+(A_\m\otimes 1 + 1\otimes A_\m ) ]   \psi_{10}  dx^\m \nn \\
  &+ {a_1\over \sqrt{2}}( \S \otimes 1+1\otimes \S ) \psi_{10} \c_2  + ( H \otimes 1 ) \psi_5 \c_3, \label{d1} \\
 \D \psi(x,2) &=  {a_1\over \sqrt{2}} [ \del_\m+( P A_\m P \otimes 1 + 1\otimes P A_\m P )] \psi_{10}  dx^\m \nn \\
 &+ {a_1\over \sqrt{2}}( P \S P \otimes 1+1\otimes P \S P ) \psi_{10} \chi_1  + (P H \otimes 1 ) \psi_5 \chi_3, \label{d2}\\
 \D \psi(x,3) & =(  \del_\m+A_\m +  P A_{\m} P ) \psi_{5}  dx^\m +{a_1\over \sqrt{2}}( 1\otimes H^{\dg} )  \psi_{10} \chi_1 
+ {a_1\over \sqrt{2}}( 1\otimes H^{\dg} P )  \psi_{10} \chi_2 \label{d3} .
\end{align}
The associated spinor-one form is written as
\begin{align}
   {\tilde \D }\psi (x,1) &= {a_1\over \sqrt{2}} ( \gamma_\m  \psi_{10} dx^\m - i c_{d} \psi_{10} \chi_2 - i c_{d} \psi_{10} \chi_3 ) , \label{tilded1}\\
   {\tilde \D }\psi (x,2) &= {a_1\over \sqrt{2}} ( \gamma_\m  \psi_{10} dx^\m - i c_{d} \psi_{10} \chi_1 - i c_{d} \psi_{10} \chi_3 ) , \\
   {\tilde \D }\psi (x,3) &= \gamma_\m \psi_{5} dx^\m - i c_{d} \psi_{5} \chi_1 - i c_{d} \psi_{5} \chi_2 .    \label{tilded3} 
\end{align}
Summarizing the above discussion, we can obtain the Dirac Lagrangian
\begin{align}
 {\Lg}_{D}&= a_1^2 \, {\rm tr} \, {\bar \psi_{10}} i \g^\m( \del_\m + 2A_\m^{\hat a} T^{\hat a} \otimes 1+1+\otimes 2 A_\m^{\hat a} T^{\hat a} )\psi_{10} \nn \\
& + {\bar \psi_{5}} i \g^\m( \del_\m+2A_\m^{\hat a} T^{\hat a} ) \psi_{5} - y'_{d} [{\bar \psi_{10}} (H + PH) \psi_{5} + {\rm h.c.}] .
 \label{Ldown}
\end{align}
Note that the coupling between $\S$ and $\psi_{10}$ disappears by the chirality condition,  
$\bar \psi_{10} \psi_{10} = 0.$ 
In particular, the explicit form of the Yukawa interactions are just those of the SM:
\begin{align}
y'_{d} \, \bar \psi_{10} (H + PH) \psi_{5} + {\rm h.c.} = 2 y'_{d}  (\bar q_{L \a}  d_{R} + \bar l_{L \a} e_{R}) H_{\rm SM}^{\a} + {\rm h.c.} ,
\end{align}
where $H_{\rm SM}^{\a} = i (\s^{2})^{\a\b} H^{*}_{\b} $, with $H^{*}_{\b} = (H^{4,5})^{*}$.
The Yukawa coupling constant is $y'_{d} = 2\alpha^{2} c_{d}$.

\vspace{18pt}

In Eq.~(\ref{Ldown}), it is clear that up-quarks are still massless. 
Thus, we consider the up-type Yukawa interactions hereafter.
Since the Dirac Lagrangian is written as an inner product form, 
the up-type Yukawa interaction requires the introduction of a completely antisymmetric fermion $\tilde \psi_{10}$,  
 which is transformed as a ${10}^*$ representation of SU(5). 
In components, $\tilde \psi_{10}$ is expressed as
\begin{align}
    \tilde \psi_{10}^{ijk}={a_2\over \sqrt{24}}\epsilon^{ijklm}(\psi_{10}^c)_{lm} ,    \label{assign1}
\end{align}
where $\epsilon^{ijklm}$ is the completely antisymmetric tensor of SU(5), and 
$\psi_{10}^c$ is the charge conjugation of $\psi_{10}$. 
We simply assign these $\tilde \psi_{10}$ and $\psi_{10}$ to each space as
\begin{align}
     \psi'(x,1)=\psi'(x,2)= \tilde \psi_{10}, ~~~ \psi'(x,3)= a_{3} \psi_{10}. \label{assign2}
\end{align}
From Eqs. (\ref{assign1}) and (\ref{assign2}),  the gauge transformation properties 
of $\psi'(x,n)$ are:
\begin{align}
\psi'^g(x,(1,2)) &= g(x,(1,2))\otimes g(x,(1,2))\otimes g(x,(1,2)) \psi'(x,(1,2)), \\
\psi'^g(x,3) &= [g(x,1)+g(x,2)] \otimes [g(x,1)+g(x,2)] \psi'(x,3). \label{gaugetrf32}
\end{align}
Then, the covariant spinor one-form in Eq.(\ref{covpsi}) is found to be
\begin{align}
 \D \psi' (x,1)&= [  \del_\m+( A_\m\otimes 1\otimes 1 
  + 1\otimes A_\m\otimes 1 +1\otimes 1\otimes A_\m ) ] \tilde \psi_{10} dx^\m  \nn \\
   &+( \S \otimes 1\otimes 1 +1\otimes \S \otimes 1
   + 1\otimes 1\otimes\S ) \tilde \psi_{10} \chi_{2}
   + ( H \otimes 1\otimes 1 ) \psi_{10} \chi_3,  \\
 \D \psi' (x,2)&= [  \del_\m+( P A_\m P \otimes 1\otimes 1 
  + 1\otimes P A_\m P \otimes 1 +1\otimes 1\otimes P A_\m P ) ] \tilde \psi_{10} dx^\m  \nn \\
   &+( P \S P \otimes 1\otimes 1 +1\otimes P \S P \otimes 1
   + 1\otimes 1\otimes P \S P  ) \tilde \psi_{10} \chi_{1}
   + ( P H \otimes 1\otimes 1 ) \psi_{10} \chi_3, \\
 \D \psi' (x,3) &=(  \del_\m+ [A_\m + P A_{\m} P] \otimes 1 + 1\otimes [A_\m + P A_{\m} P] ) \psi_{10} dx^\m \nn \\
 &+( 1\otimes 1\otimes H^{\dg} )  \tilde \psi_{10} \chi_1   +( 1\otimes1\otimes H^{\dg} P )  \tilde \psi_{10} \chi_2.
\end{align}
Similarly, the associated spinor one-form is written as
\begin{align}
  {\tilde \D }\psi'_{1} &= \gamma_\m  \tilde \psi_{10} dx^\m - i c_{u} \tilde \psi_{10} \chi_2 -i c_{u} \tilde \psi_{10} \chi_3  , \\
  {\tilde \D }\psi'_{2} &= \gamma_\m  \tilde \psi_{10} dx^\m - i c_{u} \tilde \psi_{10} \chi_1 -i c_{u} \tilde \psi_{10} \chi_3  , \\
  {\tilde \D }\psi'_{3} &= a_{3} [\gamma_\m \psi_{10} dx^\m -i c_{u} \psi_{10} \chi_1 -i c_{u} \psi_{10} \chi_2 ], 
\end{align}
where $c_{u}$ is the related up-type Yukawa coupling constant.
Accordingly, we obtain the Dirac Lagrangian for the second assignment:
\begin{align}
   {\cal L}'_{D}&= ( a_2^2+a_3^2 ) \, {\rm tr} \, {\bar \psi_{10}} i \g^\m
 ( \del_\m+ 2A_\m^{\hat a} T^{\hat a} \otimes 1+1+\otimes 2 A_\m^{\hat a} T^{\hat a} ) \psi_{10} \nn \\
 & -  \, y_{u}' [ \bar{\tilde{\psi}}_{10} ( (H+PH)  \otimes\psi_{10}) + {\rm h.c.} ], 
 \label{Lup}
\end{align}
with $y_{u}' = 2 c_{u} \a^{2}$. 
In particular, the explicit form of the Yukawa interactions are just those of the SM:
\begin{align}
y_{u}' \bar{\tilde{\psi}}_{10} ( (H+PH)  \otimes\psi_{10}) + {\rm h.c.} &=
y_{u}' {a_{2} \over \sqrt {24}} \e_{ijklm} (\bar{\psi}^{c}_{10})^{ij} ( 2 H_{\rm SM}^{k} \psi_{10}^{lm} ) + {\rm h.c.}  \\ 
&= y_{u}' { 8 a_{2} \over \sqrt {6}}  \, \bar q_{L} u_{R} \tilde H^{\rm SM}  + {\rm h.c.} ,
\end{align}
where $\tilde H^{\rm SM} = i \s^{2} H^{\rm SM \, *}$. 
Finally, summarizing the results of Eqs.~(\ref{Ldown}) and (\ref{Lup}), 
 we obtain the final form of the Dirac Lagrangian of the SU(5) GUT:
\begin{align}
  \Lg = {\bar \psi_{\rm SM}} i \gamma^\m( \del_\m+A_\m^{\hat a} T^{\hat a} ) \psi_{\rm SM} 
 - [y_{u} \, \bar q_{L} u_{R} \tilde H^{\rm SM} + y_{d} (\bar q_{L} d_{R} + \bar l_{L} e_{R}) H^{\rm SM}  + {\rm h.c.}] \,.
\end{align}
Here, $\psi_{SM}$ represents the SM fermions $q_{L}, u_{R}, d_{R}, l_{L}, e_{R},$\footnote{We normalized the kinetic term of decuplets $\psi_{10}$ as $a_1^2+a_2^2+a_3^2=1$, $[u_{R}, q_{L}, e_{R}]\to {1\over \sqrt{2}} [u_{R}, q_{L}, e_{R}]$, and $A_{\m}^{\hat a} \to {1\over \sqrt 2} A_{\m}^{\hat a}$. }
and the Yukawa couplings are found to be $y_{u} = 8 \a^{2} c_{u} a_{2} / \sqrt{6}  $ and $y_{d} = \sqrt 2 \alpha^{2} c_{d}$.
This fermionic Lagrangian is just that of the SM. 
The couplings between fermions and the heavy bosons $X_{\m}, Y_{\m}$, and $H^{c}$ 
are prohibited by the parity symmetry. 
As a result,  proton decay is forbidden at tree level.
If quantum fluctuation respects the parity symmetry, 
the process will be naturally suppressed or even forbidden completely.  
Although the process is too suppressed to detect in this case,  
other GUT groups such as SO(10), and other parity assignments, 
might allow the baryon number violating interactions.

Meanwhile, the application of the orbifold GUT mechanism to the fermion sector may be meaningful for model building. 
In early papers on the NCG \cite{Coquereaux:1990ev, Morita:1993zv, Chamseddine:1992kv, Chamseddine:1992nx}, 
the {\it ad hoc} chiral condition is usually imposed on the fermions 
in order to produce chiral Yukawa couplings. 
However, if we consider the NCG model as a model with an extra dimension, 
especially with several branes \cite{Hashimoto:1999qi}, 
these chiral fermions will generate a brane anomaly. 
This is a mirror fermion problem \cite{Maalampi:1988va}. 
By the orbifold GUT mechanism, we can eliminate the mirror fermion in proper situation. 
This enables us to implement new kind of theory in NCG: family unification in the extra dimension \cite{Babu:2002ti}. 

\subsection{Discussion}

In this subsection, we comment on an obscure point in the above construction of the Lagrangian.  
In fact, the conditions Eq.~(\ref{gaugetrf3}) and (\ref{gaugetrf32}) are seem to be {\it ad hoc} conditions.
The down-type Lagrangian (\ref{Ldown}) is also  schematically described as
\begin{align}
\Lg _{D} &= \sum_{n,m} \bar \psi_{n} [i \D_{M} \G^{M}]_{nm}  \psi_{m} \nn \\
&= 
\begin{pmatrix}
{a_{1} \over \sqrt 2} \bar \psi_{10} & {a_{1} \over \sqrt 2} \bar \psi_{10} & \bar \psi_{5}
\end{pmatrix}
\begin{pmatrix}
A_{\m} \g^{\m} & \S & H \\ P \S P & PA_{\m} P \g^{\m} & P H \\ H^{\dg} & H^{\dg} P & ([A_{\m} + PA_{\m} P] \g^{\m})
\end{pmatrix}
\begin{pmatrix}
{a_{1} \over \sqrt 2} \psi_{10 } \\ {a_{1} \over \sqrt 2} \psi_{10 } \\ \psi_{5} 
\end{pmatrix} .
\label{scheme}
\end{align}
Here, $\D_{M} \G^{M}$ is the extension of the Dirac operator with $M = 0,1,2,3,5$.
The 33 component of the matrix $([A_{\m} + PA_{\m} P] \g^{\m})$ is introduced only for fermions in this formulation. 
However, 
in the original paper on SU(5) GUT in NCG \cite{Chamseddine:1992kv, Chamseddine:1992nx}, 
the authors treated fermions as a matrix rather than a vector in Eq.~(\ref{scheme}): 
\begin{align}
\Psi = 
\begin{pmatrix}
{a_{1} \over \sqrt 2} \psi_{10 } \\ {a_{1} \over \sqrt 2} \psi_{10 } \\ \psi_{5} 
\end{pmatrix}
~~ \To ~~ 
\Psi^{IJ} = 
\begin{pmatrix}
{1\over \sqrt{6}} \psi_{10}^{ij} & 0 & {1\over \sqrt{2}} \psi_{5}^{i} \\
0 & {1\over \sqrt{6}} \psi_{10}^{ij} & {1\over \sqrt{2}} \psi_{5}^{i} \\
- {1\over \sqrt{2}} \psi_{5}^{j} & -{1\over \sqrt{2}} \psi_{5}^{j} & 0
\end{pmatrix} ,
\end{align}
where ${1\over \sqrt{2}}$, and ${1\over \sqrt 6}$ are the normalization coefficients like $a_{1,2,3}$. 
This corresponds to the situation where the fermion $\psi_{5}$ is treated as a ``link field,'' or an intrinsic differential one-form $\psi = \psi(x) \chi_{1,2}$.
In this case, we can anticipate that the undesirable gauge interaction will cancel between 
two $\psi_{5}$s, and then the orbifold GUT mechanism also works successfully without the 33 component term $([A_{\m} + PA_{\m} P] \g^{\m})$.
Otherwise, we can also solve this point by adding an additional noncommutative extra dimension (then the model would be a six-dimensional theory), such as $M^{4} \times Z_{2} \times Z_{2}$ \cite{Konisi:1998ur}. Introduction of $\psi_{5}$ in two separated points leads to proper cancellation of the gauge interaction.

\section{Conclusions}

In this paper, we applied the orbifold GUT mechanism to the SU(5) model in NCG, including the fermonic sector. 
Imposing proper parity assignments for ``constituent fields'' of bosons and fermions, 
the couplings between fermions and the heavy bosons $X_{\m}, Y_{\m}$, and $H^{c}$ 
are prohibited by the parity symmetry. 
As a result, the derived fermionic Lagrangian is just that of the SM, and proton decay is forbidden at tree level.
If quantum fluctuation respects the parity symmetry, 
the process will be naturally suppressed or even forbidden completely.  

Moreover, the application of the orbifold GUT mechanism to the fermion sector may be meaningful for 
model building. 
In early papers on the NCG \cite{Coquereaux:1990ev, Morita:1993zv, Chamseddine:1992kv, Chamseddine:1992nx}, 
the {\it ad hoc} chiral condition is usually imposed on the fermions 
in order to produce chiral Yukawa couplings. 
However, if we consider the NCG model as a model with an extra dimension, 
especially with several branes \cite{Hashimoto:1999qi}, 
these chiral fermions will generate a brane anomaly. 
This is a mirror fermion problem \cite{Maalampi:1988va}. 
By the orbifold GUT mechanism, we can eliminate the mirror fermion in the proper situation. 
This enables us to implement a new kind of theory in NCG: family unification in the extra dimension \cite{Babu:2002ti}. 

\section*{Acknowledgement}

This study is supported by the Iwanami Fujukai Foundation.


\begin{thebibliography}{10}

\bibitem{Georgi:1974sy}
H.~Georgi and S.~Glashow,
\newblock Phys.Rev.Lett. {\bf 32}, 438 (1974).

\bibitem{Pati:1974yy}
J.~C. Pati and A.~Salam,
\newblock Phys.Rev. {\bf D10}, 275 (1974).

\bibitem{Dimopoulos:1981zb}
S.~Dimopoulos and H.~Georgi,
\newblock Nucl.Phys. {\bf B193}, 150 (1981).

\bibitem{Sakai:1981gr}
N.~Sakai,
\newblock Z.Phys. {\bf C11}, 153 (1981).

\bibitem{Ramond:1979py}
P.~Ramond,
\newblock arXiv:hep-ph/9809459.

\bibitem{Wilczek:1981iz}
F.~Wilczek and A.~Zee,
\newblock Phys.Rev. {\bf D25}, 553 (1982).

\bibitem{Kugo:1983ai}
T.~Kugo and T.~Yanagida,
\newblock Phys.Lett. {\bf B134}, 313 (1984).

\bibitem{Raby:2008gh}
S.~Raby,
\newblock Eur.Phys.J. {\bf C59}, 223 (2009), arXiv:0807.4921.

\bibitem{Babu:2002ti}
K.~Babu, S.~Barr, and B.~Kyae,
\newblock Phys.Rev. {\bf D65}, 115008 (2002), arXiv:hep-ph/0202178.

\bibitem{Connes:1990qp}
A.~Connes and J.~Lott,
\newblock Nucl.Phys.Proc.Suppl. {\bf 18B}, 29 (1991).

\bibitem{Connes:1994yd}
A.~Connes, {\it Noncommutative Geometry}
\newblock (Academic Press, London 1994).

\bibitem{DuboisViolette:1988ir}
M.~Dubois-Violette, R.~Kerner, and J.~Madore,
\newblock J.Math.Phys. {\bf 31}, 316 (1990).

\bibitem{Coquereaux:1990ev}
R.~Coquereaux, G.~Esposito-Farese, and G.~Vaillant,
\newblock Nucl.Phys. {\bf B353}, 689 (1991).

\bibitem{Sitarz:1993zf}
A.~Sitarz,
\newblock Phys.Lett. {\bf B308}, 311 (1993), arXiv:hep-th/9304005.

\bibitem{Morita:1993zv}
K.~Morita and Y.~Okumura,
\newblock Prog.Theor.Phys. {\bf 91}, 959 (1994).

\bibitem{Chamseddine:1992kv}
A.~H. Chamseddine, G.~Felder, and J.~Frohlich,
\newblock Phys.Lett. {\bf B296}, 109 (1992).

\bibitem{Chamseddine:1992nx}
A.~H. Chamseddine, G.~Felder, and J.~Frohlich,
\newblock Nucl.Phys. {\bf B395}, 672 (1993), arXiv:hep-ph/9209224.

\bibitem{Morita:1993xj}
K.~Morita and Y.~Okumura,
\newblock Prog.Theor.Phys. {\bf 91}, 975 (1994).

\bibitem{Okumura:1994ck}
Y.~Okumura,
\newblock Phys.Rev. {\bf D50}, 1026 (1994), arXiv:hep-th/9402047.

\bibitem{Sogami:1996xy}
I.~Sogami,
\newblock Prog.Theor.Phys. {\bf 95}, 637 (1996).

\bibitem{Hashimoto:1999qi}
K.~Hashimoto,
\newblock Prog.Theor.Phys. {\bf 102}, 419 (1999), arXiv:hep-th/9903115.

\bibitem{Konisi:1998ur}
G.~Konisi, Z.~Maki, M.~Nakahara, and T.~Saito,
\newblock Prog.Theor.Phys. {\bf 101}, 1105 (1999), arXiv:hep-th/9812065.

\bibitem{Chamseddine:1993is}
A.~H. Chamseddine and J.~Frohlich,
\newblock Phys.Rev. {\bf D50}, 2893 (1994), arXiv:hep-th/9304023.

\bibitem{Lizzi:2000bc}
F.~Lizzi, G.~Mangano, and G.~Miele,
\newblock Mod.Phys.Lett. {\bf A16}, 1 (2001), arXiv:hep-th/0009180.

\bibitem{Sarrazin:2009ea}
M.~Sarrazin and F.~Petit,
\newblock Phys.Rev. {\bf D81}, 035014 (2010), arXiv:0903.2498.

\bibitem{Alishahiha:2001nb} 
  M.~Alishahiha,
  Phys.\ Lett.\ B {\bf 517}, 406 (2001), arXiv:hep-th/0105153.
  
\bibitem{Kawamura:2000ev}
Y.~Kawamura,
\newblock Prog.Theor.Phys. {\bf 105}, 999 (2001), arXiv:hep-ph/0012125.

\bibitem{Yang:2015gsa}
M.~J.~S. Yang,
\newblock Prog.~Theor.~Exp.~Phys. {\bf 2015}, 043B10 (2015), arXiv:1501.03888.

\bibitem{Hebecker:2002rc}
A.~Hebecker and J.~March-Russell,
\newblock Phys.Lett. {\bf B539}, 119 (2002), arXiv:hep-ph/0204037.

\bibitem{Altarelli:2001qj}
G.~Altarelli and F.~Feruglio,
\newblock Phys.Lett. {\bf B511}, 257 (2001), arXiv:hep-ph/0102301.

\bibitem{Maalampi:1988va}
J.~Maalampi and M.~Roos,
\newblock Phys.Rept. {\bf 186}, 53 (1990).

\end{thebibliography}
\end{document}